\documentclass[showpacs,preprintnumbers,amsmath,amssymb]{revtex4}

\usepackage{graphicx}
\usepackage{dcolumn}
\usepackage{bm}

\begin{document}



\title{On the gauge and BRST invariance of the chiral QED with
Faddeevian anomaly}

\author{Anisur Rahaman}
\affiliation{Durgapur Govt. College, Durgapur - 713214, Burdwan,
West Bengal, India}
\email{anisur.rahman@saha.ac.in}
\author{Safia Yasmin}%
\affiliation{Indas Mahavidyalaya, Bankura - 722205, West Bengal,
India}
\author{Sahazada Aziz}
\affiliation{Burdwan University, Burdwan-713104, West Bengal,
India}



\date{\today}

\begin{abstract}
Chiral Schwinger model with the Faddeevian anomaly is considered.
It is found that imposing a chiral constraint this model can be
expressed in terms
 of chiral boson. The model when expressed in terms of chiral boson
 remains anomalous and the Gauss law of which gives anomalous Poisson
 brackets between itself. In spite of that a systematic BRST quantization is
 possible. The Wess-Zumino term corresponding to this theory appears automatically
 during the process of quantization. A gauge invariant reformulation of this model
  is also constructed.
Unlike the former one gauge invariance is done here without any
extension of phase space. This gauge invariant version  maps onto
the vector Schwinger model. The gauge invariant version of the
chiral Schwinger model for $a=2$ has a massive field with
identical mass however gauge invariant version obtained here does
not map on to that.
\end{abstract}

 \maketitle

\section{Introduction}
Symmetry plays a fundamental role in physics.
 Some times symmetry of
a given theory may be broken and that has a profound consequences.
Gauge symmetry of a theory is of particular interest in this
context. Absence of gauge symmetry invites anomaly in a theory.
There have been considerable efforts in the understanding of
anomaly in quantum field theory \cite{JR, ROT1, ROT2, FLO, FAD1,
FAD2, SHATAS0, KH, PM, MG, AR1, AR2, AR3, AR4}. The studies of
chiral Schwinger model and anomalous Schwinger model \cite{AR1}
are worth mentionable in this respect. It is the anomaly that
removed the long suffering of chiral Schwinger model from
non-unitarity. Credit went to Jackiw and Rajaraman - those who
offered a consistent analysis of this model in a gauge
non-invariant manner \cite{JR}. However a gauge invariant version
is always favorable because of its increased symmetry.  This work
is an illustration on gauge as well as BRST invariance of chiral
Schwinger mode with Faddeevian \cite{FAD1, FAD2} type of anomaly.

In terms of constraint \cite{DIR}, a gauge invariant theory is
defined as a theory with first class constraint and the presence
of second class constraints  indicate the breaking of this
invariance. The conversion mechanism of second class constraints
into a first class was introduced initially by Faddeev and
Shatashvili in \cite{FAD2}. The formalism was extended further by
Batalin, Fradkin and Vilkovisky \cite{FR1, FR2, FR3, FR4, FR5} and
became amenable for obtaining BRST invariant effective action. It
is known so as BFV formalism. There have been attempts for this
conversion in different approaches too. The approaches basically
fall into two independent classes. In one class extension of phase
space through the introduction of auxiliary fields is required
\cite{FR1, FR2, FR3, FR4, FR5, FR6, FR7}. The other class however
does not require this extension \cite{MR1, MR2}.

Study of free chiral boson \cite{FLO, SIG, BEL, IMB, JMF} as well
as gauged chiral boson \cite{KH, PM, MG, BEL} are very interesting
in connection with the restoration of gauge invariance because of
its peculiar constraint structure. To be precise chiral
constraint$^1$ \footnotetext [1] {Chiral constraint is a relation
between the momenta and coordinate of the chiral boson.
Mathematically it is $T(x)=\pi_\phi-\Phi'\approx 0$. The Poisson
bracket between $T(x)$ and  $T{y}$ is $[T(x), T(y)]=
-2\delta'(x-y).$ } shows non vanishing Poisson bracket with
itself. Gauge invariant reformulation of free chiral boson and
gauged chiral boson are considered by several authors in different
time \cite{FR5, FR6, FR7, MR1, MR2, SG, MIAO}. It is known that
two independent version of gauged chiral boson are available in
the literature. The oldest one is the version proposed by Jackiw
and Rajaraman \cite{JR}. We should mention here that Hagen
initially gave the chiral generation of Schwinger model
\cite{HAG}. The model however failed to maintain the unitarity.
Jackiw and Rajaraman saved the model introducing anomaly \cite{JR}
within it and gave a consistent hamiltonian description of that.
Mitra suggested an alternative gauge non-invariant version of
gauged chiral boson \cite{PM, MG}. The anomaly of which
corresponds to Faddeevian type \cite{FAD1, FAD2, SHATAS0} where
gauss law constraint gave nonvanishing Poisson bracket among
itself. The model attracted several attention because of this
special type of costraint structure.

BRST invariant reformulation of Jackiw-Rajaraman version of
    chiral Schwinger model is done
   in \cite{FR5, SG}. However the gauge (BRST) invariant reformulation is
   lacking for the chiral Schwinger model
   where anomaly is Faddeevian like. It would be worthy to have
   a systematic development where
   gauge invariance gets restored
and the Wess-Zumino term comes out automatically during the
process. With this in view and also as a pedagogical illustration
of the BVF formalism effort has been made to obtain a BRST
invariant effective action of this model. The work will certainly
demonstrate the power of BFV formalism once more. This new study
would be instrumental for future studies towards unitarity and
renormalization of this model. Gauge invariant reformulation of
this model also carried out in its usual phase space using Mitra-
Rajaraman prescription in order to have a better feeling about the
difference between gauge invariance in the usual phase space and
the extended phase space. In \cite{SHATAS}, Shatashvili considered
the non-Abelian version of gauge invariant chiral Schwinger model
where he observed an special feature in connection with the
reduction of interacting degrees of freedom of this model for
$a=2$. That work also showed that the mass term in that was
identical to the mass term of the  model considered here for that
specified value of $a$. At the first sight one may think that
these two models are identical but that is not the case.  To get a
clear picture we compare our present development with the work of
Shatashvili \cite{SHATAS}.

The paper is organized as follows. Sec. II, contains a brief
review of the model in connection with the bosonization of the
fermionic version of chiral Schwinger model and imposition of a
chiral constraint to express it in terms of chiral boson.  In Sec.
III, a brief introduction of the the BFV formalism is given and
then it is applied to this model to obtain the BRST invariant
reformulation of that. Mitra-Rajaraman  prescription is used in
Sec. IV to obtain a gauge invariant reformulation of the same
model. In Sec. V, a comparison is made between the result obtain
in Sec. IV, and and the known gauge invariant version of the usual
chiral Schwinger model for a=2. Sec. VI contains a brief
discussion over the work.

\section{Bosonization of fermionic model and imposition of chiral constraints}
Chiral Schwinger model is described by the following generating
functional
\begin{equation}
Z[A] = \int d\psi d\bar\psi e^{\int d^2x{\cal L}_f},
\end{equation}
with
\begin{eqnarray}
{\cal L}_f &=& \bar\psi\gamma^\mu[i\partial_\mu +
e\sqrt\pi A_\mu(1-\gamma_5)]\psi \nonumber \\
 &=& \bar\psi_R\gamma^\mu i\partial_\mu\psi_R +
\bar\psi_L \gamma^\mu(i\partial_\mu + 2e\sqrt\pi A_\mu)\psi_L
.\end{eqnarray} The right handed fermion remains uncoupled in this
type of chiral interaction. So integration over this right handed
part leads to field independent counter part which can be absorbed
within the normalization. Integration over left handed fermion
leads to
\begin{equation}
Z[A] = exp[{\frac{ie^2}{2}}\int d^2x A_\mu[M^{\mu\nu} -
(\partial^\mu +\tilde\partial^\mu) \frac{1}{\Box} (\partial^\nu +
\tilde
\partial^\nu)]A_\nu].
\end{equation} $M_{\mu\nu} = ag_{\mu\nu}$,
for Jackiw-Rajaraman regularization where the parameter $a$
represents the regularization ambiguity and
$M_{\mu\nu} = \left(\begin{array}{cc} 1 & -1\\
-1 & -3 \\
\end{array}\right)\delta(x-y)$, for an alternative version proposed in \cite{PM, MG}.
Writing down the generating functional in terms of the auxiliary
field $\phi(x)$ it turns out to the following
\begin{equation}
Z[A] = \int d\phi e^{i\int d^2x {\cal L}_B},
\end{equation}
with
\begin{eqnarray}
{\cal L}_B &=& \frac{1}{2} (\partial_\mu\phi)(\partial^\mu\phi) +
e(g^{\mu\nu} - \epsilon^{\mu\nu)}\partial_\mu\phi A_\nu +
\frac{1}{2} e^2 A_\mu M^{\mu\nu}A_\nu \nonumber\\
&=& \frac{1}{2}(\dot\phi^2 - \phi'^2) + e(\dot\phi + \phi')(A_0 -
A_1) + \frac{1}{2}e^2(A_0^2 - 2 A_0A_1 -3 A_1^2) .\label{NE}
\end{eqnarray} Here $\epsilon^{01}=-\epsilon_{01}=1$ and the
Minkowski metric $g^{\mu\nu}=diag\left(1,-1\right)$.

Equation (\ref{NE}) was initially found in \cite{PM} where Mitra
termed it as chiral Schwinger model with Faddeevian
regularization. In \cite{PM}, we find that the Gauss law
constraint of this theory is
\begin{equation}
G = \pi_1' + e(\pi_\phi + \phi')
\end{equation}
It is found there that the Poisson bracket between G(x) and G(y)
is
\begin{equation} [G(x), G(y])] = 2\delta(x-y)'. \label{POIS}
\end{equation}
This Poisson (\ref{POIS}) was found to gave the vanishing
contribution for the usual chiral Schwinger model \cite{JR}.
Faddeev initially noticed that anomaly made Poisson bracket
between $G(x)$ and $G(y)$ nonzero \cite{FAD1, FAD2}. The
constraint became second class itself and gauge invariance was
lost. He, however, argued that it would be possible to quantize
the theory but in this situation system may posses more degrees of
freedom.

From the standard definition, the momentum corresponding to the
field $\phi$ is found out to be
\begin{equation}
\pi_\phi = \dot\phi + e(A_0 - A_1) .
\end{equation}
The following Legendre transformation
\begin{equation}
H_B = \int d^2x [\pi_\phi\dot\phi - {\cal L}_B],
\end{equation}
leads to the hamiltonian density
\begin{eqnarray}
{\cal H}_B &=& \frac{1}{2}[\pi_\phi - e(A_0 - A_1)]^2
+ \frac{1}{2}\phi'^2 - 2e\phi'(A_0 - A_1) \nonumber \\
&-& \frac{1}{2}e^2(A_0^2 -2 A_0A_1 -3A_1^2).
\end{eqnarray}
In order to suppress one chirality at this stage we impose the
chiral constraint
\begin{equation} \omega(x) = \pi_\phi(x) -
\phi'(x) \approx 0. \end{equation}
It is a second class constraint
itself since
\begin{equation}
[\omega(x), \omega(y)] = -2\delta'(x-y). \end{equation} After
imposing the constraint $\omega(x) \approx  0$, into the
generating functional we arrived at the following
\begin{eqnarray}
Z_{CH} &=& \int d\phi d\pi_\phi \delta(\pi_\phi - \phi')
\sqrt{det[\omega, \omega]}e^{ i\int d^2x(\pi_\phi\dot\phi - {\cal
H}_B)}
\nonumber \\
&=&\int d\phi e^{i\int d^2x{\cal L}_{CH}}, \end{eqnarray} with
\begin{equation}
{\cal L}_{CH} = \dot\phi\phi' -\phi'^2 + 2e(A_0 - A_1)\phi' + 2e^2
A_1^2.\label{LCH}
\end{equation}
We  obtained the gauged lagrangian for chiral boson from the
bosonized lagrangian with Faddeevian regularization \cite{PM} just
by imposing the chiral constraint in its phase space. Harada in
\cite{KH}, obtained the same type of result for the usual chiral
Schwinger model with one parameter class of regularization
proposed by Jackiw and Rajaraman \cite{JR}. The lagrangian
(\ref{LCH}) can be thought of as the gauged version of chiral
boson described by Floreanini and Jackiw \cite{FLO}. The
constraint analysis and the phase space structure corresponding to
this model is available in \cite{MG}.  In ref. \cite{MG}, we found
that the theory (\ref{LCH}) describes a massive boson through the
equation
\begin{equation}
[\Box + 4e^2]A_1 = 0 \label{BEQ}
\end{equation} with square of the mass $m^2 = 4e^2$. Equation (\ref{BEQ}) was
interpreted there as the photon acquired mass and the fermion got
confined.

\section{BRST invariant reformulation using BFV formalism}
Before we apply BFV formalism on this model it would be useful for
the reader to give a brief introduction of the formalism. BRST
invariance essentially means to enlarge the Hilbert space of a
gauge theory in order to restore the symmetry of a gauge fixed
action in that enlarged space.  It is very effective when one
tries to study the renormalization property of a theory. One
generally exploit the BRST symmetry instead of exploiting the
original gauge symmetry. The discovery of this symmetry raised the
ghost field to a prominent position. It mixes the ghost with the
other  fields of the theory and therefore all the fields including
the ghosts can be regarded as a different components of a single
geometrical object.

The combined formalism of Batalin, Fradkin and Vilkovisky
\cite{FR1, FR2, FR3, FR4, FR5} for quantization of a system is
based on the idea that a system with second class constraint can
be made effectively first class in the extended phase space which
finally helps to find  BRST invariant effective action. The field
needed for this conversion ultimately turns out into the
Wess-Zumino scalar with the proper choice of gauge condition, as
pointed out by Fugiwara, Igarashi and Kubo \cite{FR3}. What
follows next is  a brief description of the general BFV formalism
for obtaining a BRST invariant action.

Let us consider a canonical hamiltonian described by  the
canonical pairs $(p_i, q^i),  i=1,2.......N$. The pairs are
subjected to a set of constraints $\Omega_a \approx 0,
a=1,2.......n$, and it is assumed that the constraints satisfy the
following algebra \cite{FR5, FR6}.
\begin{equation}
[\Omega_a, \Omega_b] = i\Omega_cU^c_{ab},
\end{equation}
\begin{equation}
[H_c, \Omega_a] = i\Omega_bV^b_c,
\end{equation}
then $n$ no of additional condition $\Phi_a \approx 0$ with
$det[\Phi_a, \Omega_b] \neq 0$ have to be imposed in order to
single out the physical degrees of freedom. The  constraints
$\Omega_a \approx 0$ and $\Phi_a \approx 0$, together with
hamiltonian equation of motion is obtained from the action
\begin{equation}
S = \int dt[p_i\dot q^i - H_c(p_i,q^i) - \lambda^a\Omega_a +
\pi_a\Phi^a],
\end{equation}
where $\lambda^a$ and $\pi_a$ are Lagrange multiplier fields and
these two satisfy the relation $[\lambda^a, \pi_b] = i\delta^a_b$.

Now introducing one pair of canonical ghost field
$(C^a,\bar{P}_a)$ and one pair of canonical anti-ghost field
$(P^a,\bar{C}_a)$ for each pair of constraints an equivalence can
be made to the initial theory with constraints in the reduced
phase space. So the quantum theory can be described by the
partition function where the action \cite{FR1, FR2, FR3, FR4, FR5,
FR6, FR7} in its numerator will be
\begin{equation}
S_{qf} = \int dt[p_i\dot q^i + \pi^a\dot\lambda_a + \bar{P}^a\dot
C_a + \bar{C}^a \dot P_a- H_{BRST} + i[Q, \psi]].
\end{equation}
$H_{BRST}$ is the minimal hamiltonian \cite{FR1, FR2}as termed by
Batalin and Fradkin, is defined by
\begin{equation}
H_{BRST} = H_c + \bar{P}_aV^a_bC^b.
\end{equation}
The BRST charge $Q$ and the fermionic gauge fixing function $\psi$
are respectively given by \cite{FR5, FR6, FR7}
\begin{equation}
Q = C^a\omega_a - \frac{1}{2} C^bC_cU^c_{ab}\bar{P}^a + P^a\pi_a,
\end{equation}
\begin{equation}
\psi = \bar{C}_c\chi^a + \bar{P}^a\lambda^a,
\end{equation}
where $\chi_a$'s are expressed through the gauge fixing condition
\begin{equation}
\Phi_a =\dot\lambda_a
 + \chi_a
\end{equation}

Let us now concentrate on the BRST invariant reformulation of the
lagrangian (\ref{LCH}). In order to do that we need to know the
constraint structure of the theory. The details of which is
available in \cite{PM, MG}. Here we are giving the relevant
portion as required for our purpose. In \cite{MG}, we find that
the momenta corresponding to the fields $A_0$, $A_1$ and $\phi$
are.
\begin{equation}
\pi_{\phi}=\phi',
\end{equation}
\begin{equation}
\pi_{1} = \dot{A}_{1}-{A}'_{0},
\end{equation}
\begin{equation}
\pi_{0}=0.
\end{equation}
It is known that  $\pi_0 = 0$ and $\pi_\phi = \phi'$ are  the
primary constraints of the theory.

The effective hamiltonian follows from the equations of motion is
\begin{equation}
H_P=\int dx[ {\cal
H}_{C}+u\pi_{0}+v\left(\pi\phi-{\phi}'\right)]\label{EFFH},
\end{equation}
where
\begin{equation}
{\cal
H}_{C}=\frac{1}{2}\pi_{1}^2+\pi_{1}A'_{0}+{\phi}^2-2e\left(A_{0}-A_{1}\right){\phi}'-2e^2A_{1}^2
\end{equation}
Here $u$ and $v$ are two required lagrange multipliers. The
preservation of the constraints leads to two other constraints
\begin{equation}
G ={\pi}'_{1} + 2e{\phi}'\approx 0 , \label{GAUSS}
\end{equation}
\begin{equation}
-2e^2(A_1 + A_0)'\approx 0. \label{FINAL}
\end{equation}
The multipliers $u$ and $v$ are found out to be
\begin{equation}
u= -(\pi_1 + A_0'),
\end{equation}
\begin{equation}
v=\phi -e(A_0 -A_1).
\end{equation}
Therefore, the theory under consideration contains four
constraints in its phase space. Precisely, the constraints are
\begin{equation}
\omega_1=\pi_{\phi}-\phi'\approx 0, \label{CONS1}
\end{equation}
\begin{equation}
\omega_2 = \pi_{0}\approx 0,\label{CONS2}
\end{equation}
\begin{equation}
\omega_3 = {\pi}'_{1} + 2e{\phi}'\approx 0, \label{CONS3}
\end{equation}
\begin{equation}
\omega_4 = -2e^2(A_1 + A_0)' \approx 0.\label{CONS4}
\end{equation}
These four constraints form a second class set and the closures of
the constrains with respect to the hamiltonian (\ref{EFFH}) are
given by
\begin{equation}
\dot\omega_1 = \omega'_1,
\end{equation}
\begin{equation}
\dot\omega_2 = \omega_3 - \omega'_2 +  e\omega_1,
\end{equation}
\begin{equation}
\dot\omega_3 = \omega_4 - e\omega'_1,
\end{equation}
\begin{equation}
\dot\omega_4 = 2e^2\omega'_2.
\end{equation}

To obtain a BRST invariant reformulation we need to convert the
second class set of constraints into a first class set. With this
in view, we introduce four auxiliary fields $\psi$ , $\eta$,
$\pi_\psi$ and $\pi_\eta$ and fields are such that they satisfy
the following canonical condition
\begin{equation}
\left[\eta(x),\pi_\eta(y)\right]=\delta(x-y),
\end{equation}
\begin{equation}
\left[\psi(x),\pi_\psi(y)\right]=\delta(x-y).
\end{equation}
The fields used here are known as Batalin-Fradkin (BF) fields. The
constraints (\ref{CONS1}), (\ref{CONS2}), (\ref{CONS3}) and
(\ref{CONS4}), with some suitable linear combination of the BF
fields get converted into first class set as follows
\begin{equation}
\tilde{\omega}_1 =\pi_\phi- {\phi}'+\pi_\psi+{\psi}',
\end{equation}
\begin{equation}
\tilde{\omega}_2=\pi_0-\pi_\eta,
\end{equation}
\begin{equation}
\tilde{\omega}_3= -2e{\psi}'+2e{\phi}'+{\pi}'_1-{\pi}'_\eta,
\end{equation}
\begin{equation}
\tilde{\omega}_4=-2e^2(A_0 +A_1)' -2e^2{\eta}'.
\end{equation}
The above four  first class constraints will be found consistent
with the first class hamiltonian
 if these new first class set satisfy the same closures as their
ancestor did with the hamiltonian (\ref{EFFH}). Precisely, the
conditions are
\begin{equation}
\dot{\tilde\omega}_1 =\tilde \omega'_1 \label{TDC1},
\end{equation}
\begin{equation}
\dot{\tilde\omega}_2 = \tilde\omega_3 - \tilde\omega'_2 +
e\tilde\omega_1\label{TDC2},
\end{equation}
\begin{equation}
\dot{\tilde\omega}_3 = \tilde\omega_4 - e\tilde\omega'_1
\label{TDC3},
\end{equation}
\begin{equation}
\dot{\tilde\omega}_4 = 2e^2\tilde\omega'_2\label{TDC4}.
\end{equation}
First class hamiltonian is obtained by the appropriate insertion
of the BF fields within the hamiltonian (\ref{EFFH}) and it is
given by  $\tilde{H}$=$H_{P}$+$H_{BF}$. Here $H_{BF}$ is a
polynomial of  $\psi$ , $\eta$, $\pi_\psi$ and $\pi_\eta$ that
extend the phase space respecting the closures (\ref{TDC1}),
(\ref{TDC2}), (\ref{TDC3}) and (\ref{TDC4}). We find that $H_{BF}$
for this system will be
\begin{equation}
H_{BF}=\int dx[
-2e\eta{\psi}'+e(\pi_\psi+{\psi}')\eta+\frac{1}{2}(\pi_\eta^2+\pi_\psi^2
+{\psi'}^2)].
\end{equation}

We now introduce four pairs of ghost $(C_i,\bar{P}^i)$ and four
pairs of anti-ghost $(P_i,\bar{C}^i)$ fields.  Four pairs of
multiplier fields $(N^i,B_i)$ are also needed. The pairs satisfy
the following canonical relations
\begin{eqnarray}
\left[C_i,\bar{P}^j\right]=\left[P^i,\bar{C}_j\right]=\left[N^i,B_j\right]=i\delta^i_j\delta(x-y)
, \hspace{2cm} i=1,2,3,4
\end{eqnarray}
From the definition we can write  BRST invariant hamiltonian
\begin{equation}
H_U=H_{BRST} -i[Q,\psi],
\end{equation}
where $H_U$ is the unitarizing hamiltonian, $Q$ is the BRST charge
and $\psi$'s are the gauge fixing functions. Note that the BRST
charge $Q$ is a nilpotent operator and it satisfies the equation
\begin{equation}
Q^2=[Q,Q]=0.
\end{equation}
The definition of $Q$ in this formalism is
\begin{equation}
Q=\int(B_iP^i+C_i\tilde{\omega}^i)dx ,
\end{equation}
and the definition of gauge fixing function $\psi$ is
\begin{equation}
\psi =\int (\bar{C_i}X^i+P_iN^i)dx.
\end{equation}
The BRST invariant hamiltonian for the theory with which we are
dealing with is
\begin{eqnarray}
H_{BRST}&=& H_{P}+H_{BF}+\int
dx(-\bar{P}_1C'_1+\bar{P}_3C_2+\bar{P}_2C'_2+e\bar{P}_1C_2+\bar{P}_4C_3
\nonumber \\
&-& e\bar{P}'_1C_3+2e^2\bar{P}'_2C_4).
\end{eqnarray}
It would be helpful to write down the generating functional that
ultimately leads to an effective action with the elimination of
some fields by Gaussian integration.  The generating functional
reads
\begin{equation}
Z=\int [D\mu]e^{iS} \label{GEN}.
\end{equation}
Here the expression of $S$ is
\begin{eqnarray}
S &=& \int d^2x[\pi_\phi\dot{\phi} + \pi_1\dot{A}_1+
\pi_0\dot{A}_0 + \pi_\psi\dot{\psi}+ \pi_\eta\dot{\eta} +
\bar{P}_i\dot{C}^i+\bar{C}_i\dot{P}^{i} +B_i\dot{N}^i \nonumber
\\
&-& H_U], \label{ACT}
\end{eqnarray}
where $[D\mu]$ is the Liouville measure in the extended phase
space. We are now in a position  to fix up the gauge conditions
\begin{equation}
\chi_1=\pi_\phi-\phi', \label{GF1}
\end{equation}
\begin{equation}
\chi_2=-\dot{N}^2+A_0,\label{GF2}
\end{equation}
\begin{equation}
\chi_3=\frac{B_3}{2}-A_1',\label{GF3}
\end{equation}
\begin{equation}
\chi_4=\pi_\eta-\dot{N}^4.\label{GF4}.
\end{equation}
When we substitute the simplified form of $H_{final}$ obtained
 after plugging the gauge fixing conditions (\ref{GF1}), (\ref{GF2}), (\ref{GF3})
and (\ref{GF4}) in the action (\ref{ACT}), we get the explicit
expression of $S$:
\begin{eqnarray}
S&=&\int d^2x[\pi_\phi\dot\phi+
\pi_\psi\dot\psi+\pi_\eta\dot{\eta} + \pi_1\dot {A}_1+ \pi_0\dot
A_0+\bar{P}_i\dot{C}^i+\bar{C}_i\dot{P}^{i} +B_i\dot{N}^i\nonumber\\
&-& (\frac{\pi^2}{2}+\pi A_0'+e\phi'(A_0-A_1)+2e^2A^2_1
-\pi_0(\pi_1+A_0')+\pi_\phi\phi' \nonumber\\
&-& e\pi_\phi(A_0-A_1)
-e\eta\psi'+e\pi_{\psi}\eta+\frac{1}{2}(\pi^2_\eta+\pi^2_\psi+{\psi'^2})
+B_i\chi^i+\tilde{\omega}_iN^i \nonumber\\
&-& \bar{P}_iP^i -\bar{P}_1C_1'+\bar{P}_3C_2+\bar{P}_2C_2'
+e\bar{P}_1C_2+\bar{P}_4C_3-e\bar{P}'_1C_3+2e^2\bar{P}_2'C_4 \nonumber\\
  &-&
  C_3\bar{C}''
-\bar{C}_2\dot{P}^2-\bar{C}_4\dot{P}^4+2e^2C^4\bar{C}_4 -
2C'\bar{C}_1 - e^2C^2\bar{C}_2)].
\end{eqnarray}
Here $i$ runs from $1$ to $4$. Our next task is to  simplify
equation (\ref{GEN}) through the elimination of some fields and
that will lead us to our desired result. A careful look reveals
that here exists a simplification
\begin{equation}
\int d^2x(B_iN^i+\bar{C}_i\dot{P}^i) =-i[Q,\int
d^2x(\bar{C}_i\dot{N}^i)]
\end{equation}
with be Legendre transformation $B^i  \rightarrow B^i+\dot{N}^i$.
However the simplification corresponding to $i=1$ suffices in this
situation. More simplification follows from the elimination of the
fields $\pi_0$, $\pi_1$, $\pi_\eta$, $B_1, B_2, B_4, A_0, N^1$,
$N^2, N^4, P_1, \bar{P}^1, P_2$, $\bar{P}^2, P_4, \bar{P}^4, P_1,
\bar{P}^1$, $C_1, \bar{C}^1, C_2$ and $\bar{C}^2$ by Gaussian
integration. Ultimately we reach to a very simplified form of the
generating functional (\ref{GEN}) that contains the following
effective action in its numerator.
\begin{eqnarray}
S_{eff}&=&\int
d^2x(\dot{\phi}\phi'-\phi'^2+2e^2A_1^2-\psi'^2-\dot{\psi}\psi'
+\frac{1}{2}(\dot{A_1}-A_0')^2 \nonumber\\ &+& 2e\phi'(A_0-A_1)-
2e\psi'(A_1+A_0) + \partial_\mu B A^\mu + \frac{1}{2}\alpha B^2
\nonumber \\
 &+& \partial_{\mu}\bar C \partial_\mu C
\label{FACT}
\end{eqnarray}
We have used few redefinition  of fields, e.g., $N_3=A_0$ and
$P^3$=$\dot{C}_3$ to reach to the result (\ref{FACT}). Since after
elimination there is no other $B$'s and $C$'s except $B_3$ and
$C_3$ we are free to read them as $B$ and $C$. It is now time to
check the invariance of the action (\ref{FACT}). A little algebra
shows that the action is invariant under the transformation
\begin{eqnarray}
\delta A_1 &=& -\lambda C', \delta A_0 = \delta N_3 = \lambda\dot C \nonumber \\
\delta \psi &=& \lambda C, \delta \bar C = \lambda B, \delta C = 0
\end{eqnarray}
It is to be mentioned that the fields satisfy the following
Euler-Lagrange equation
\begin{equation}
\partial_-\phi - \partial_+\psi - 2eA_1= 0 \label{WLEQ}
\end{equation}
We can identify easily the Wess-Zumino term for this theory which
is
\begin{equation}
{\cal L}_{wz} = -\dot\psi\psi' - \psi'^2 - 2e\psi'(A_0 + A_1)
\end{equation}
   It is interesting  to see this automatic appearance of this Wess-Zumino
term  during the process of obtaining the BRST invariant action.
One point we should mention here that the choice of gauge
condition is very crucial. One may miss to get Wess-Zumino term
otherwise.
\section{Gauge invariant reformulation without extending the phase space}
The formalism of making a theory gauge invariant by the reduction
of the number of second class constraint was first developed by
Mitra and Rajaraman \cite{MR1, MR2}. The formalism strictly
depends on the constraint structure of the
 theory. Depending on the constraint structure of the theory different
gauge invariant version is possible for a particular theory. No
extension of phase space is needed in this formalism. So the
physical contents of all the gauge invariant actions remains the
same. In \cite{MR1, MR2}, the authors gave a reasonably general
theory relating to a large class of systems with second class
constraints to corresponding class of gauge invariant systems
having the same dynamical content. A gauge theory in a generalized
sense means a theory with some first class constraints. To covert
it into an equivalent second class system is  well known. One
generally fix the gauge, i.e., impose a suitable number of gauge
fixing conditions. These gauge fixing  conditions together with
the original first class set of constraint form a second class set
and the theory gets  converted into an equivalent second class
system. An inverse procedure is suggested in \cite{MR1, MR2} where
a formalism is developed for construction of a gauge invariant
system equivalent to a given second class theory. The authors
argued there as follows. If a
 dynamical system possess $2n$ constraints and  the constraints all
together form a second class set and if $n$ of these constraints
are found to have mutually vanishing Poisson brackets then these
$n$ constraints can be used as gauge generator of the gauge
invariant reformulation. The remaining $n$ constraints may be
thought of as the gauge fixing condition. The hamiltonian needs
 the required modification accordingly.
So in \cite{MR1, MR2} the authors suggested to reduce half of the
constraint from a second class  set of constraint retaining the
first class set only in order to get the gauge invariant
reformulation. The obtained gauge invariant theory can be treated
in the similar way as any standard gauge invariant theory is
treated. What follows next is the the application of the formalism
in the presently considered mode.

 To apply this formalism in a model it is essential to know  the constraint
 structure of that theory. In our case which is already given in Sec. III. We have
 seen there that the
 phase space of the
model described by the lagrangian (\ref{LCH}) contains
 four constraints. In Sec III, those constraints are given in (\ref{CONS1}), (\ref{CONS2}),
(\ref{CONS3}) and (\ref{CONS4}).
Note that the combination $\omega_2\approx 0$ and $\omega_3
\approx 0$ form a first class set. If we retain only these two
constraints as stated above, following the suggestion available in
\cite{MR1, MR2}, we require a modification of the hamiltonian
density of the second class system (\ref{EFFH}) in the following
manner in order to get a first class system.
\begin{eqnarray}
{\cal H}&=&
\frac{1}{2}\pi_1^2+\pi_1A_0'-e(A_0-A_1)\phi'+2e^2A_1^2+
\pi_{\phi}\phi'-e\pi_{\phi}(A_0-A_1)  \nonumber
\\
&+& e(\pi_{\phi}-\phi')(A_0 + A_1) +
\frac{1}{2}(\pi_{\phi}-\phi')^2 + u\pi_0. \label{MODH}
\end{eqnarray}
The modification certainly keeps the physical contents of the
theory intact. This modified hamiltonian density (\ref{MODH})
contains only the two first class constraints $\omega_2 \approx 0$
and $\omega_3 \approx 0$.  The equation of motion with respect to
the hamiltonian (\ref{MODH}) are found out as follows
\begin{equation}
\dot{\phi} = [\phi, {\cal H}] = \pi_{\phi}+ 2eA_1 ,\label{EQ1}
\end{equation}
\begin{equation}
\dot{A_0}= [A_0, {\cal H}]= -u\label{EQ2},
\end{equation}
\begin{equation}
\dot{A_1} = [A_1, {\cal H}] = \pi_1\label{EQ3}.
\end{equation}
A straightforward calculation leads to the lagrangian density
corresponding to the first class theory with which we are
interested in.
\begin{eqnarray}
{\cal L}_1
&=&\pi_{\phi}\dot{\phi}+\pi_{1}\dot{A_1}+\pi_{0}\dot{A_0}-[\frac{\pi_1^2}{2}
+\pi_1A_0'+2eA_1\pi_{\phi}
+\pi_{\phi}\phi'-2eA_0\phi'\nonumber \\
&+&\frac{1}{2}(\pi_{\phi}-\phi')^2+u\pi_0+2e^2A_1^2].
\end{eqnarray}
After a little algebra the lagrangian density acquires a very
simplified form
\begin{equation}
{\cal L}_2=\frac{1}{2}(\dot{\phi}^2-\phi'^2) -2e(A_1\dot{\phi}
-A_0\phi')+\frac{1}{2}(\dot{A_1}-A_0')^2 \label{FLAG}
\end{equation}
The lagrangian density(\ref{FLAG}), is consistent with the
hamiltonian density (\ref{MODH}), and the equations of motion
(\ref{EQ1}), (\ref{EQ2}) and (\ref{EQ3}). To see whether the
lagrangian density (\ref{FLAG}) stems out from the modified
hamiltonian density (\ref{MODH}) contains only the two first class
(\ref{CONS2}) and (\ref{CONS3}) in its phase space let us
calculate the momenta corresponding to the field $A_0$
\begin{equation}
\pi_0= \frac{\partial {\cal L}_2}{\partial\dot A_0} = 0.
\label{CC1}
\end{equation}
 It gives back the primary constraint (\ref{CONS3})
and the preservation of this once again gives the Gauss law
constraint
\begin{equation}
G ={\pi}'_{1} + 2e{\phi}'\approx 0.\label{CC2}
\end{equation}
No other constraints come out from the preservation of
(\ref{CC2}). These two first class constraints help us to
construct the gauge transformation generator. The generator is
given by
\begin{equation}
{\cal G} = \int dx(\lambda_1\omega_1 + \lambda_2\omega_2),
\label{GENERATOR}
\end{equation}
Here $\lambda_1$ and $\lambda_2$ are two arbitrary parameters. The
transformations evolved out of the generator (\ref{GENERATOR}) for
the fields $\phi$, $A_1$ and $A_0$ respectively are
\begin{equation}
\delta \phi =0, \delta A_1 = -\lambda_1',  \delta A_0 =
-\lambda_2. \label{TRAN}
\end{equation}
A little algebra shows that under the transformation (\ref{TRAN}),
the lagrangian (\ref{FLAG}) remains invariant provided the
parameter satisfy the relation
\begin{equation}
\lambda_2=\dot{\lambda_1}.
\end{equation}
A note worthy thing is that this transformation is equivalent to
the transformation $A_\mu \rightarrow A_\mu +
\frac{1}{2e}\partial_\mu\lambda$. There is some thing interesting
that we must mention here. The first class lagrangian that comes
out from our investigation is the bosonized lagrangian of the well
known vector Schwinger model \cite{SCH, LOW}. Here coupling
strength is $2e$. It does not come as a great surprise because the
theoretical spectrum of the model under consideration is identical
to the vector Schwinger model. To be precise, both the models
contain the massive boson with mass $m = 2e$.

We have mentioned earlier that the gauge invariant reformulation
follows from this prescription depends crucially on the constraint
structure of the model.  There are other possibilities to get
first class set of constraints from the set of constraints
(\ref{CONS1}), (\ref{CONS2}), (\ref{CONS3}) and (\ref{CONS4}).
However that possibilities fail to give consistent first class
theories.

\section{Comparison of the result obtained in Sec. IV with the gauge
invariant chiral Schwinger model for $a=2$} Let us compare our
result with the work of the Shatashvili \cite{SHATAS} because
seeing their apparent similarities at a first glance one may think
that these two results are identical. But a careful look revels
that this is not so. In his work Shatashvili considered the
non-Abelian gauge invariant version of the chiral Schwinger model
and showed that the interacting degrees of freedom gets reduced if
the choice $a=2$ is made. For $a=2$, the mass term of
Shatashvili's model become identical to our model but there lies a
basic difference which we would like to address. Here we consider
the gauge invariant Abelian bosonized version of that model
\cite{MIYAKE} because this version would be compatible for
comparison with our work. Unlike the non-Abelian version the
Abelian version it is exactly solvable too.

It is described by the lagrangian density
\begin{equation}
{\cal L} = \frac{1}{2} (\partial_\mu\phi)(\partial^\mu\phi) +
e(g^{\mu\nu} - \epsilon^{\mu\nu)}\partial_\mu\phi A_\nu +
\frac{1}{2} a e^2 A_\mu A^\mu - \frac{1}{4} F_{\mu\nu}F^{\mu\nu}+
L_{WESS},
\end{equation} \label{NEW}
where $L_{WESS}$ is given by
\begin{equation}
{\cal L}_{WESS} = \frac{1}{2}(a-1)
(\partial_\mu\eta)(\partial^\mu\eta) + e[(a-1)g^{\mu\nu} +
\epsilon^{\mu\nu}]\partial_\mu\eta A_\nu .
\end{equation}
The lagrangian is invariant under the gauge transformation $A_\mu
\rightarrow A_\mu + \frac{1}{e}\partial_\mu\Lambda$, $\phi
\rightarrow \phi + \Lambda$, $\eta \rightarrow \eta -\Lambda$. The
momenta corresponding to the fields $A_0$, $A_1$ and $\phi$ and
$\eta$ are
\begin{equation}
\pi_{\phi}=\dot\phi+e(A_0 - A_1),\label{A1}
\end{equation}
\begin{equation}
\pi_{1} = \dot{A}_{1}-{A}'_{0}, \label{A2}
\end{equation}
\begin{equation}
\pi_{0}=0, \label{A3}
\end{equation}
\begin{equation}
\pi_{\eta}=(a-1)\dot\eta + e[(a-1)A_0 + A_1].\label{A4}
\end{equation}
Equation (\ref{A1}), (\ref{A2}) and (\ref{A3}) are independent of
the parameter $a$. The choice
 $a=2$ brings  change only in the equation (\ref{A4}) and with that choice that turns into
\begin{equation}
\pi_{\eta}=\dot\eta + e(A_0 + A_1).
\end{equation}
A straightforward calculation shows that the canonical Hamiltonian
density for the model with $a=2$ is
\begin{eqnarray}
{\cal H}_c &=& \frac{1}{2}[\pi_1^2 + \pi_\phi^2 + \phi'^2] -
eA_1(\pi_\phi - \phi')
+ 2e^2A_1^2 + \frac{1}{2}[\pi_\eta^2 + \eta^2] - eA_1(\pi_\eta + \eta')\nonumber \\
&-& A_0(\pi_1' + e[(\pi_\phi-\phi') - (\pi_\eta +
\eta')].\label{CSA}\end{eqnarray} The phase space of the model
contains the following two constraints \cite{MIYAKE}
\begin{equation}
\Omega_1 = \pi_0\approx 0, \label{C1}
\end{equation}
\begin{equation}
\Omega_2 = \pi_1' +e(\pi_\phi-\phi')- e(\pi_\eta + \eta'). \approx
0 \label{C2}
\end{equation}
The constraint (\ref{C2}) appears as a secondary constraint in
order to preserve the constraint (\ref{C1}).
 The two constraints are first class.
 The first class constraints shows a clear indication of reduction
 of degrees of freedom because to quantize the theory two gauge
 fixing conditions are to be needed.
 Bosonized version of Vector Schwinger model (\ref{FLAG}), appeared out as
 the gauge invariant version of chiral
 Schwinger model with Faddeevian anomaly in Sec. IV,  contains the
 following two constraint
\begin{equation}
\omega_{VS1}= \pi_0\approx 0,
\end{equation}
\begin{equation}
\omega_{VS2}= \pi_1' + 2e\phi'\approx 0.
 \end{equation}
The Hamiltonian density of this bosonized version of vector
Schwinger  model (\ref{FLAG}) comes out to be
\begin{equation}
{\cal H}_{VS}= \frac{1}{2}(\pi_1^2 + \pi_\phi^2 + \phi'^2) +
\pi_1A_0'+ 2e(A_1\pi_\phi - A_0\phi')\label{VS}.
\end{equation}
It is true that both the models are gauge invariant and the
massive fields which comes out form (\ref{VS}) and (\ref{CSA})
looks almost identical. Square of the mass of the boson in each
case is $m^2 = 4e^2$. However
 the Hamiltonian (\ref{CSA}) cannot be made free from Wess-Zumino
 field $\eta$ using the  constraints (\ref{C1}) and (\ref{C2}) and the
 constraints (\ref{C1}) and (\ref{C2}) also do not map on to the constraints of
 the vector Schwinger model. On
 the contrary the Gauge invariant version as obtained in (\ref{FLAG}),
 using Mitra-Rajaraman prescription, does not contain this type of field.
 Here gauge invariance is resulted in the usual phase space.

\section{Discussion}
Gauge invariant reformulation of chiral Schwinger with Faddeevian
anomaly has been carried out in two different directions. In the
first case BFV prescription \cite{FR1, FR2, FR3, FR4, FR5} is
followed which needs an extension of phase space. The process
certainly keeps the physical contents of the theory intact. The
fields needed for the extension keep themselves allocated in the
un-physical sector of the theory. In this prescription we not only
get a BRST invariant effective action but also appropriate
Wess-Zumino term appears automatically during the process. In the
second approach Mitra-Rajaraman prescription \cite{MR1, MR2} is
followed to obtain a gauge invariant action. In this situation we
have to be restricted on the gauge invariance only because the
formalism developed till now is not adequate to obtain BRST
invariant action.  In spite of the existence of more than one
possibilities only  a particular possibility leads to a gauge
invariant action there. Surprisingly, the other possibilities fail
to do so. Only that possibility has explored to obtain gauge
invariant reformulation which renders a very interesting result.
The gauge invariant model that comes out is the lagrangian of well
known vector Schwinger model \cite{SCH, LOW} and gauge invariance
of which is obvious. It is conclusively shown here too. It is true
that the gauge non-invariant version of this model under
consideration too contains a massive boson
 like  vector Schwinger model
\cite{SCH, LOW}. We have already mentioned it. But the explicit
mapping of this model onto the vector Schwinger model is a new and
novel result. The counting of degrees of freedom also found to be
consistent. It would be interesting to investigate how a
particular Faddeevian regularized version of the chiral Schwinger
model maps onto the vector Schwinger model in its gauge invariant
version. We compare the gauge invariant lagrangian obtained in
Sec. IV with the gauge invariant version of the Abelian chiral
Schwinger model setting $a=2$ in (\ref{FLAG}. Both the model is
gauge invariant and contains a massive field with the same mass .
But for the former one gauge invariance has occurred in its usual
phase space whereas for the later it does occur in the extended
phase space.


\end{document}